\documentclass[fleqn,usenatbib]{mnras}

\usepackage{xcolor}
\usepackage[export]{adjustbox}
\usepackage{flushend}
\usepackage[utf8]{inputenc}
\usepackage[normalem]{ulem}
\usepackage{comment}
\usepackage{url}

\usepackage[compact]{titlesec}  
\titlespacing{\section}{0pt}{12pt}{7pt}
\titlespacing{\subsection}{0pt}{9pt}{4pt}

\newcommand{\frb}{FRB20201124A}
\newcommand{\rGC}{FRB20200120E}
\newcommand{\rthree}{FRB20180916B}
\newcommand{\rone}{FRB20121102A}

\newcommand{\dmu}{pc\,cm$^{-3}$}

\usepackage[T1]{fontenc}
\usepackage[normalem]{ulem}

\DeclareRobustCommand{\VAN}[3]{#2}
\let\VANthebibliography\thebibliography
\def\thebibliography{\DeclareRobustCommand{\VAN}[3]{##3}\VANthebibliography}

\usepackage{graphicx}	
\usepackage{amsmath}	
\usepackage{amssymb}	
\usepackage{newtxtext,newtxmath}

\title[Scintillation timescale of FRB~20201124A]{Scintillation timescale measurement of the highly active FRB~20201124A}

\author[Main et al.]{R.~A.~Main$^1\thanks{Email: \href{mailto:ramain@mpifr-bonn.mpg.de}{ramain@mpifr-bonn.mpg.de}}$, 
G.~H.~Hilmarsson$^{1}$,
V.~R.~Marthi$^{2}$,
L.~G.~Spitler$^{1}$,
R.~S.~Wharton$^{3}$,
S.~Bethapudi$^{1}$,
\newauthor 
D.~Z.~Li$^{4}$, 
H.-H.~Lin$^{5, 6}$ \\
$^{1}$Max-Planck-Institut f{\"u}r Radioastronomie, Auf dem H{\"u}gel 69, D-53121 Bonn, Germany \\
$^{2}$National Centre for Radio Astrophysics, Tata Institute of Fundamental Research, Post Bag 3, Ganeshkhind, Pune - 411 007, India \\
$^{3}$NASA Postdoctoral Program Fellow, Jet Propulsion Laboratory, California Institute of Technology, Pasadena, CA 91109, USA \\
$^{4}$Cahill Center for Astronomy and Astrophysics, MC 249-17 California Institute of Technology, Pasadena CA 91125, USA \\
$^{5}$Institute of Astronomy and Astrophysics, Academia Sinica, Astronomy-Mathematics Building, No. 1, Sec. 4, Roosevelt Road, Taipei 10617, Taiwan \\
$^{6}$Canadian Institute for Theoretical Astrophysics, 60 St. George Street, Toronto, ON M5S 3H8, Canada \\
}

\date{Accepted XXX. Received YYY; in original form ZZZ}

\pubyear{2021}

\begin{document}
\label{firstpage}
\pagerange{\pageref{firstpage}--\pageref{lastpage}}
\maketitle

\begin{abstract}
Scintillation of compact radio sources results from the interference between images caused by multipath propagation, and probes the intervening scattering plasma and the velocities of the emitting source and scattering screen.  
In FRB20201124A, a repeating fast radio burst (FRB) which entered a period of extreme activity, we obtained many burst detections in observations at the upgraded Giant Metrewave Radio Telescope (uGMRT)
and the Effelsberg 100\,m Radio Telescope. 
Bursts nearby in time show similar scintillation patterns, and we measure a scintillation timescale of $14.3\pm1.2$\,min and $7\pm2$\,min at Effelsberg (1370\,MHz) and uGMRT (650\,MHz), respectively, by correlating burst pair spectra.  
The scintillation bandwidth scaled to 1\,GHz is $0.5\pm0.1$\,MHz, and
the inferred scintillation velocity at Effelsberg is $V_{\mathrm{ISS}}\approx (59\pm7) \sqrt{d_{l}/2\,\rm{kpc}}~{\rm km~s}^{-1}$, higher than Earth's velocity for any screen beyond a lens distance of $d_{l} \gtrsim 400\,$pc. 
From the measured scintillation bandwidth, FRB20201124A has comparatively lower scattering than nearby pulsars, and is underscattered by a factor of $\sim 30$ or $\sim 1200$ compared to the NE2001 and YMW16 model predictions respectively.  This underscattering, together with the measured scintillation velocity are consistent with a scattering screen more nearby the Earth at $d_{l} \sim 400\,$pc, rather than at 2\,kpc spiral arm which NE2001 predicts to be the dominant source of scattering.  With future measurements, the distance, geometry, and velocity of the scattering screen could be obtained through modelling of the annual variation in $V_{\rm ISS}$, or through inter-station time delays or interferometric observations. Scintillation/scattering measurements of FRBs could help improve Galactic electron density models, particularly in the Galactic halo or at high Galactic latitudes.
\end{abstract}




\section{INTRODUCTION}

Fast Radio Bursts (FRBs) are highly dispersed, micro- to millisecond duration radio bursts of extragalactic origin.  Their dispersion measure (DM) contains the contribution of all free electrons along the line of sight, including their host galaxy, the intergalactic medium (IGM), and the Milky Way (MW).  

Along with the DM, FRBs often exhibit scintillation and scattering, effects of multipath propagation arising from inhomogeneities in the electron density along the line of sight (LOS).  These are not expected to have large contributions from the IGM, and are likely dominated by the local environment, the host galaxy, and the MW.  Many FRBs show a scintillation pattern in frequency, using which one can estimate the time delays due to multipath propagation.  As extragalactic point sources, FRBs could be used to constrain and improve models of scattering in the MW, including the halo \citep{ocker+21}.  In some cases scattering in the host and scintillation from the MW are seen simultaneously, in which case the interplay between the two can constrain the location of scattering in the host \citep{masui+15}.  

While most FRBs have been detected only once \citep{chime2021_frbsample}, repeating FRBs allow for many measurements unevenly sampled over time.  In {\rone}, this has revealed a dynamic environment showing extreme variations in the rotation measure \citep{michilli+18, hilmarsson+21}, and DM variations of $\sim 1$\dmu{} per year.  In {\rGC}, the scintillation pattern of two bursts separated by 4.3 minutes was found to partially correlate \citep{nimmo+21b}, implying that they are seen within the scintillation timescale $t_{\rm s} \gtrsim 4\,$min.  With more closely spaced bursts, one could measure the scintillation timescale for repeating FRBs, gaining additional measurements to constrain the intervening material, and potentially the emission properties and velocities of FRBs themselves - such analysis is the main focus of this work.

In March 2021, {\frb} entered a period of heightened activity \citep{chimeatel}, with burst rates of $\sim 5/$hour.  Subsequently, the source was localized to arcsecond precision \citep{askapatel, lawatel, whartonatel, marcoteatel}.  
In this paper, 
we measure a scintillation timescale of {\frb} at Effelsberg and at the uGMRT on separate days by correlating the spectra of closely spaced burst pairs.  This paper is the second of a series of papers stemming from uGMRT observations of {\frb} during its period of high activity.
The burst detection pipeline, and analysis of burst properties with the uGMRT is presented in \citet{marthi+21}, henceforth referred to as P-I, while the detection and polarization properties of Effelsberg bursts is presented in \citet{hilmarsson_Eff}.  
P-III will describe the uGMRT localization of \frb{} and the coincident persistent radio source (Wharton et al. in prep).
Section~\ref{sec:data} describes our observations, burst detections, and data reduction; Section~\ref{sec:scintillation} describes our measurements of scintillation parameters; while Section~\ref{sec:discussion} discusses analysis of derived quantities, as well as implications and potential future uses of scintillation analysis in repeating FRBs.  In Section~\ref{sec:conclusion}, we present our conclusions.

\section{DATA}
\label{sec:data}

\subsection{Observations}\label{sec:observations}

We use data taken with both the Effelsberg 100\,m telescope and the uGMRT.  Here we include a brief summary of the data and observation parameters, which are summarized in more detail in \citet{hilmarsson_Eff} and P-I.

The uGMRT observed on 2021-April-5, with a total exposure time of 3~hours towards the coordinates of the preliminary ASKAP localization \citep{askapAtelCoarse}, in order to localize the source.  Since a precise localization was not yet available, the uGMRT observation recorded the incoherent array (IA) beam \citep{uGMRTpaper}, where the antenna voltages are transformed to total intensities before being co-added, and the field of view is identical to the primary beam of a single uGMRT dish rather than the phased-array beam.
 Data were taken in band 4, in the total intensity mode at 550-750\,MHz with $655.36~{\rm \mu s}$ time resolution, and 2048 channels (97.65625 kHz channel resolution).  Effelsberg L-band observations were carried out on 2021-April-9, lasting 4 hours.  In addition to standardly recorded Effelsberg search-mode data, baseband data of the full 1210-1520\,MHz band were recorded with the Pulsar Fast Fourier Transform Spectrometer backend \citep{barr+13}.  The uGMRT and Effelsberg observations led to the detection of 48 and 20 bursts, respectively, and details of the detection pipeline and burst properties are included in P-I and \citet{hilmarsson_Eff}.

\subsection{Extracting Burst Spectra}
\label{sec:spectra}

Channels corrupted by radio-frequency interference (RFI) were manually identified and masked.  In addition to corrupted channels, the uGMRT data show variable broadband RFI spikes at DM=0,
which when dedispersed could corrupt our spectra, which would be detrimental to our scintillation measurements. To mitigate this, the frequency-averaged time series was subtracted from the data before dedispersion.

All data were dedispersed to 411.0 \dmu{}, the value identified by a visual inspection of the brightest burst (details in P-I).  At Effelsberg, two seconds of baseband data surrounding each burst were coherently dedispersed using custom software, while the uGMRT data could only be incoherently dedispersed.  Effelsberg data were produced with 4096 channels across the band by directly using a real fast Fourier Transform (FFT), corresponding to $78.125\,$kHz resolution (and discarding the zero frequency).  Only data in the range 1270-1470\,MHz were kept, to avoid the sharp drop in sensitivity at the band edges.

\begin{figure*}
    \centering
    \includegraphics[width=0.45\textwidth,scale=1.0, trim=0.0cm 1.5cm 0.0cm 2.0cm, clip=true]{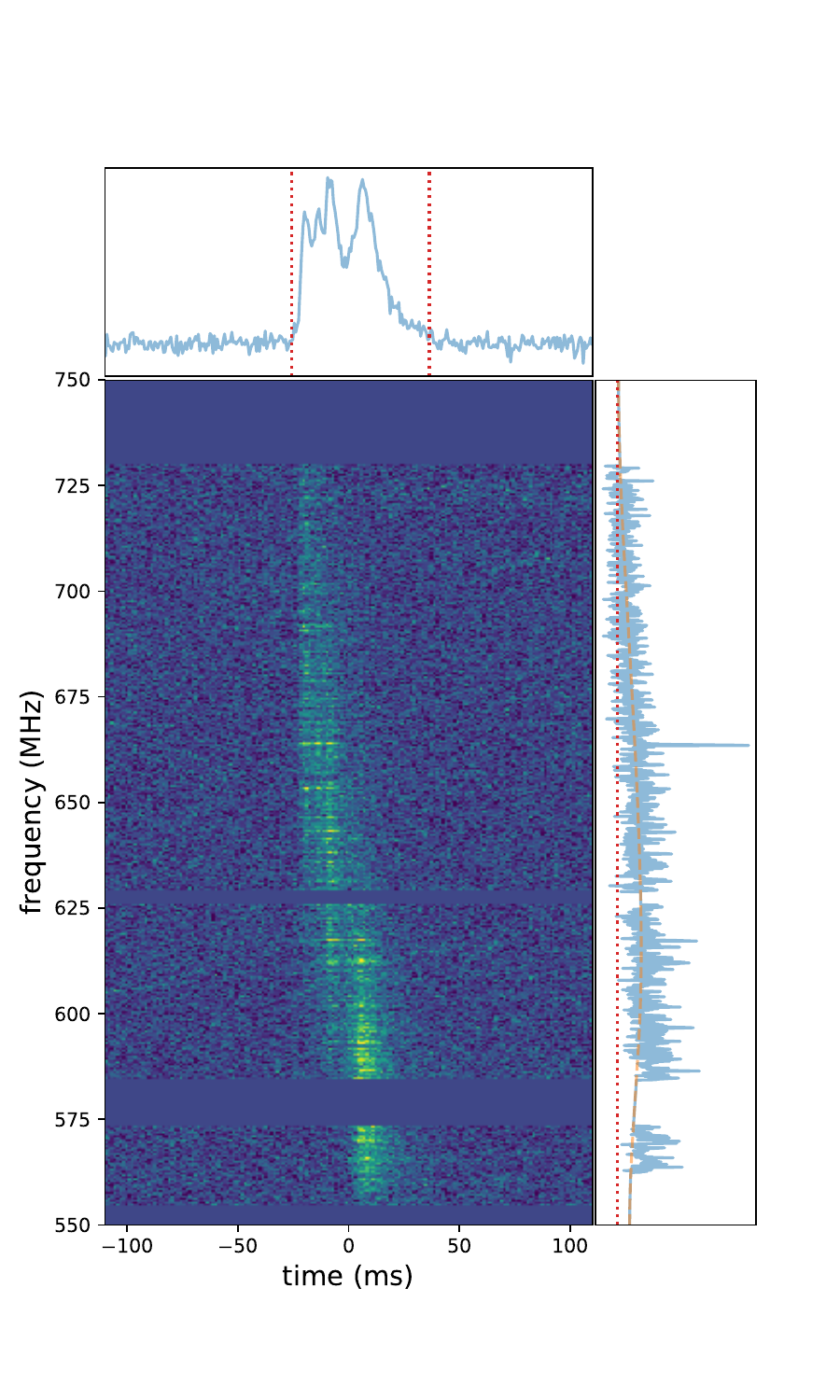}
    \includegraphics[width=0.45\textwidth,scale=1.0, trim=0.0cm 1.5cm 0.0cm 2.0cm, clip=true]{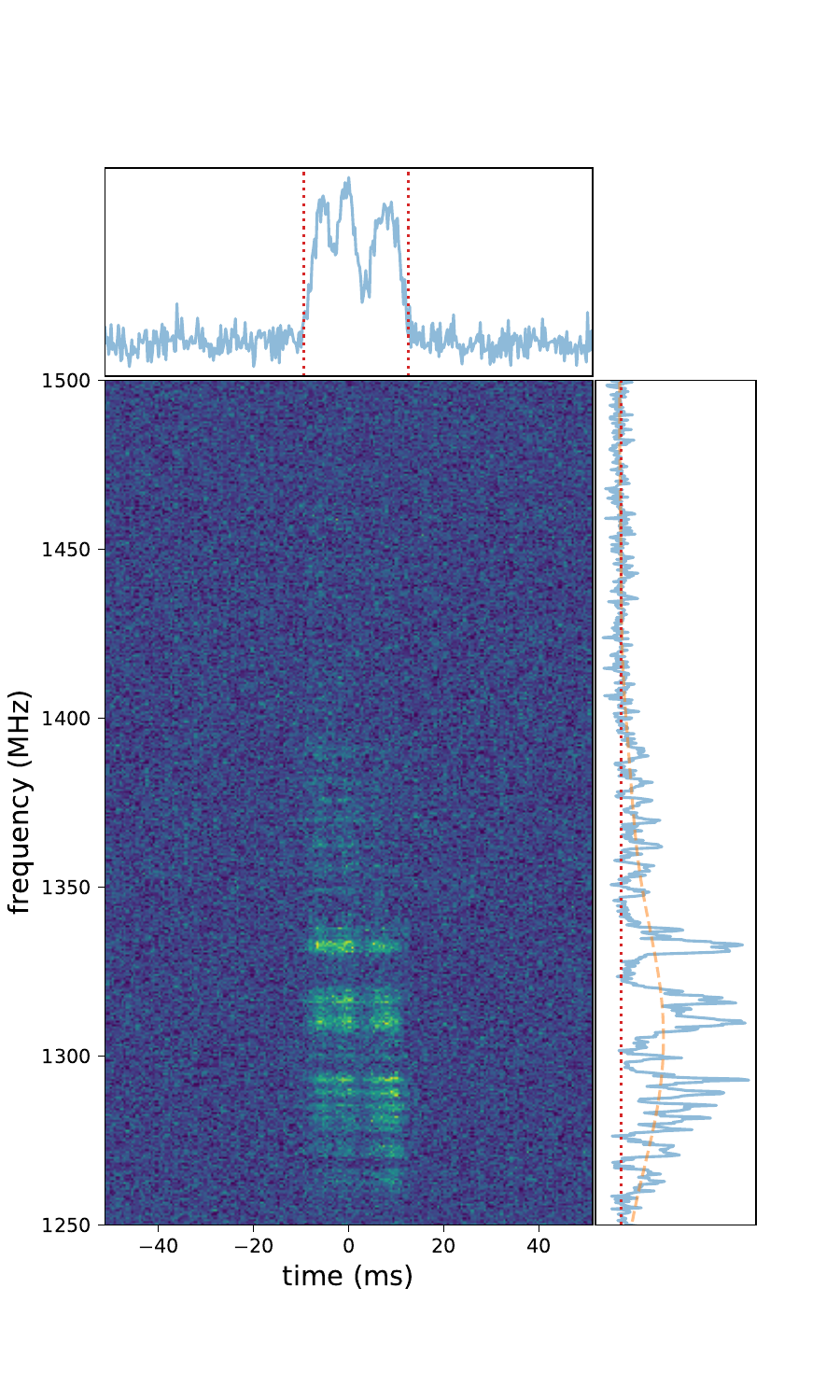}
    \vspace{-3mm}
    \caption{Diagnostic plots showing how spectra are created, with examples from uGMRT \textit{(left)} and Effelsberg \textit{(right)}. Images show the dynamic spectrum of each burst.  Top panels show the frequency averaged profile in blue, while the red dotted lines enclose the on-window containing the burst. Right panels show the spectrum (blue), and the spectrum smoothed by a Gaussian filter with a width of 1/8 of the total bandwidth. Red lines are at $0.5 \times$ the RMS noise of the background; the ``on'' region of each spectrum is determined as the channels where the smoothed spectrum is above the red line.
    }
    \label{fig:diagnostics}
\end{figure*}

Extracting the spectra of our bursts had an additional complication of individual bursts not covering the full band.  First, we divided by the background level of each channel in a region preceding each burst ($-100$--$50\,$ms ), and subtracted 1.  
An on-window in time was chosen as the contiguous window where the burst profile S/N was greater than 5.
A smoothed version of each spectrum was created (smoothed by a Gaussian filter with a width 1/8 of the band), and regions above our threshold of $0.5\sigma$ were taken as the ``on'' region (where $\sigma$ is the RMS of the background).  Our final spectrum of each burst was taken as the measured spectrum in our on-region, divided by the smoothed spectrum, to remove the broad intrinsic variations of the bursts.  
We show examples of our spectra measurements in Figure \ref{fig:diagnostics}.

\begin{figure}
    \centering
    \includegraphics[width=0.95\columnwidth,scale=1.0, trim=0.0cm 1.0cm 0.0cm 1.0cm, clip=true]{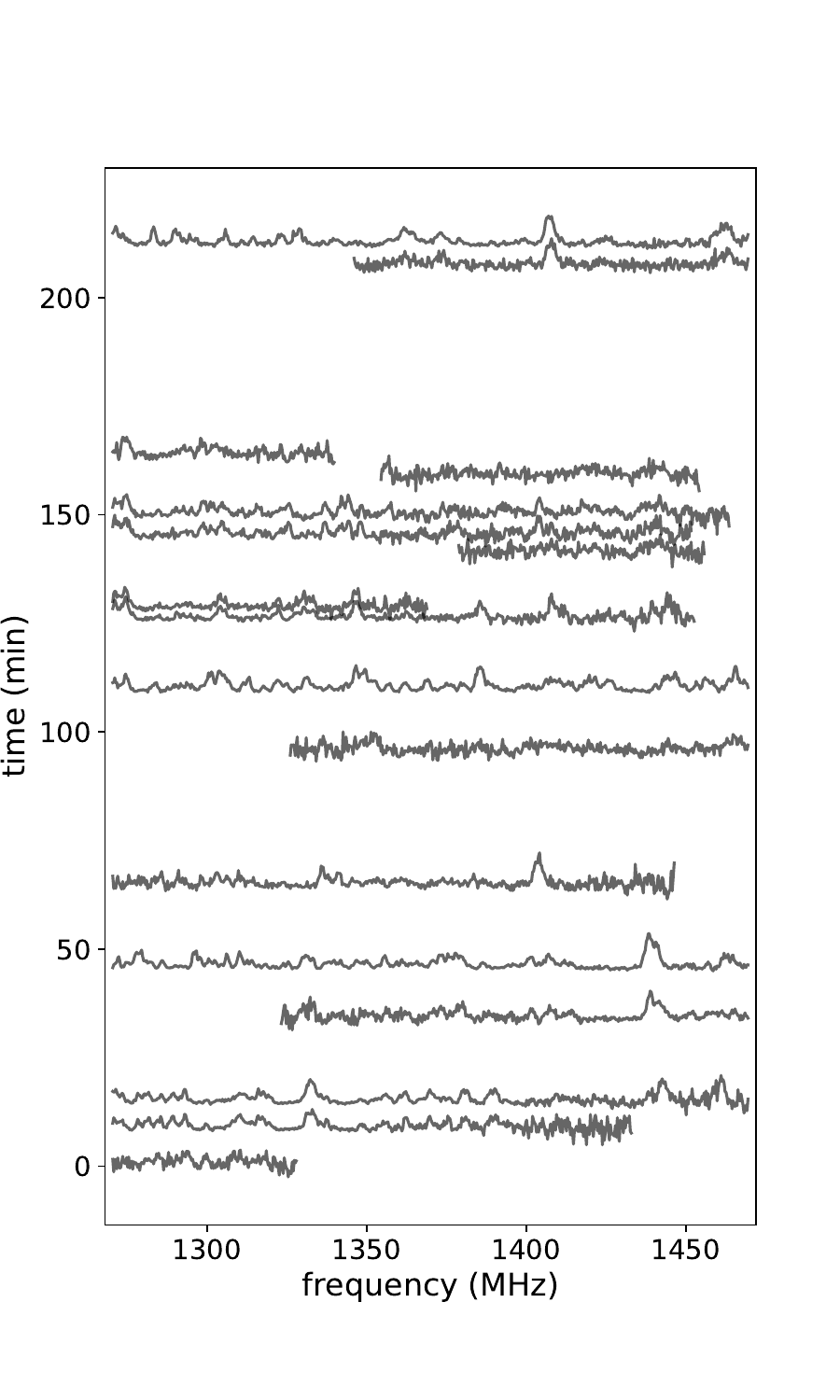} \\
    \includegraphics[width=0.95\columnwidth,scale=1.0, trim=1.0cm 0.0cm 1.0cm 1.0cm, clip=true]{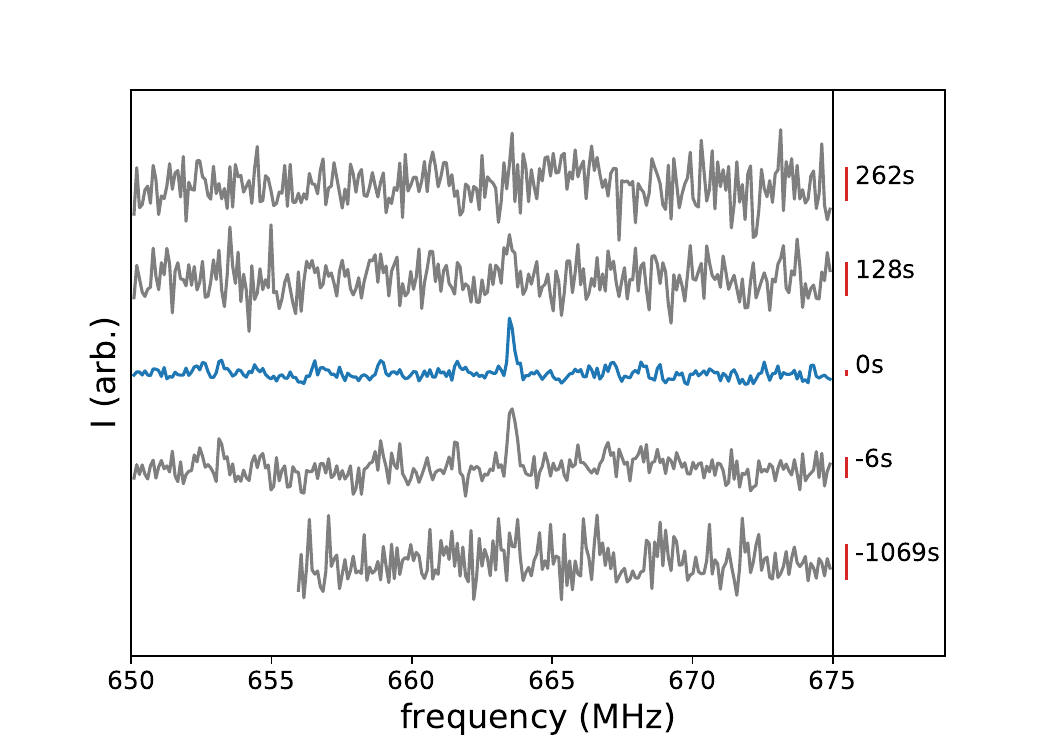}
    \caption{\textit{Top:} Spectra of all Effelsberg bursts vs. time in the observation. The frequency range is restricted to the channels where the burst is significant, as described in Section~\ref{sec:spectra}.   \textit{Bottom:} Spectra of uGMRT bursts, cropped in frequency, near the brightest pulse (\textit{blue}), used as the reference burst in computing the scintillation timescale. The red vertical bar denotes the rms noise of each burst }
    \label{fig:dynamic_spectra}
\end{figure}

\section{MEASURING SCINTILLATION PARAMETERS}
\label{sec:scintillation}

As described above in Section~\ref{sec:spectra}, we extracted the spectrum of each burst. Using the time of arrival of each burst, we plot the time ordered spectra $I(t_{i}, f)$ in Figure~\ref{fig:dynamic_spectra}, effectively an irregularly spaced version of the ``dynamic spectrum'' which is the primary observable to study scintillating pulsars.  At Effelsberg, the scintillation bandwidth is clearly visible over $\sim$\,MHz scales in frequency, and one can see that nearby bursts have highly similar spectral structures.  At uGMRT, the scintillation is much finer over frequency, seemingly only a few bins, and individual scintles are difficult to see above the noise.  However, our brightest burst has an obvious scintillation maximum at $\sim 664\,$MHz, which is apparent in the two most nearby bursts, and is not apparent in bursts more separated in time.

\subsection{Scintillation Bandwidth}
\label{sec:scint_bandwidth}

The scintillation bandwidth is typically defined as the $1/e$ scale of the frequency auto-correlation function (ACF) of a spectrum.  For the uGMRT, we restrict our measurement solely to the brightest burst, as it has more than twice the signal-to-noise (S/N) of any other single burst.  We split the full 200\,MHz band into four evenly spaced sub-bands of 50\,MHz, compute the ACF in each, and fit it with the form of a decaying exponential $r(\nu) = A e^{-\nu / \nu_{s}}$ to recover the scintillation bandwidth $\nu_{s}$.  As the scintillation features at Effelsberg are much larger, we restrict ourselves to only the four bursts which extend across the full band, and fit the scintillation bandwidth in two 100\, MHz subbands.  

\begin{figure}
    \centering
    \includegraphics[width=1.0\columnwidth,scale=1.0, trim=0.0cm 0.0cm 0.0cm 0.0cm, clip=true]{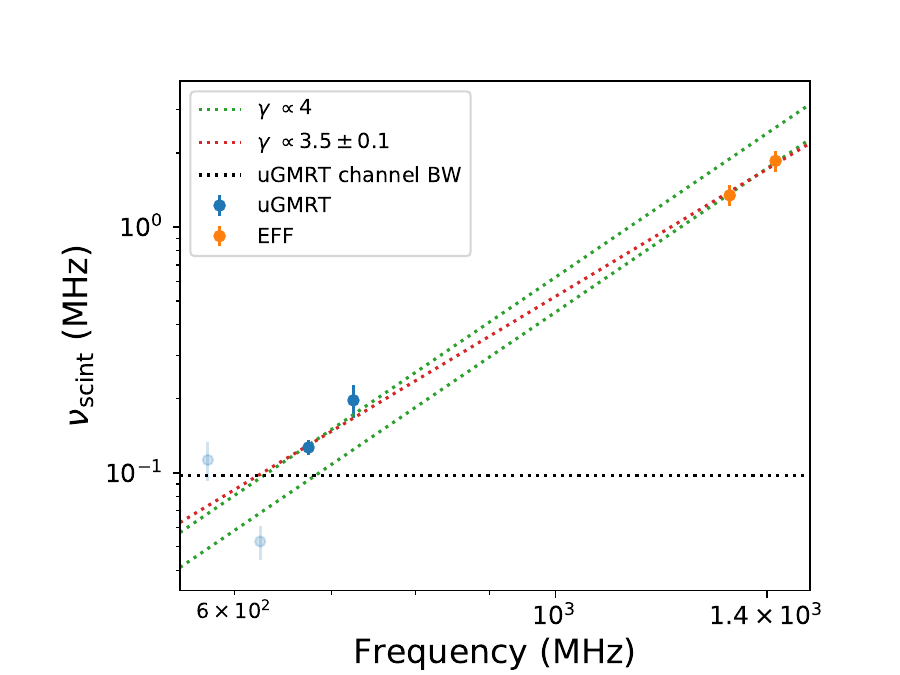} \\
    \vspace{-3mm}
    \caption{Measured scintillation bandwidth across frequency.  The red dotted line shows the best-fit power-law, while the green lines are power-laws with an index of $-4$, shown for reference. The black dotted line shows the channel bandwidth at uGMRT. The bottom two uGMRT points (shown in pale blue) have unresolved scintillation, and were excluded from the fit.}
    \label{fig:scintbandwidth}
\end{figure}

The scintillation bandwidths are shown in Figure \ref{fig:scintbandwidth}.  In the bottom two uGMRT subbands, the scintillation bandwidth is lower than the channel bandwidth and cannot be fit reliably  - these points are excluded from the following fits.  We fit the scintillation bandwidths with a power-law $\nu_{s} = a \nu^{\gamma}$, finding a best fit index of $\gamma = 3.5 \pm 0.1$, lower than the typical expectations of $\gamma=4.0$ 
or $\gamma=4.4$ for a Kolmogorov spectrum. The error is the formal statistical error -
we do caution that the uGMRT points are barely resolved, and may be biased slightly high, resulting in a shallower best fit index. 
The scintillation bandwidth extrapolated to 1\,GHz is $0.5\pm0.1$\,MHz, where the error is conservatively estimated by extrapolating from either the uGMRT or Effelsberg bandwidths to $1\,$GHz using  a $\gamma=4.0$ powerlaw.

\subsection{Scintillation Timescale}
\label{sec:scint_timescale}

In addition to measuring the scintillation bandwidths, the fact that nearby bursts shows similar spectral features allows us the possibility of measuring a scintillation timescale.  For bursts irregularly spaced in time, the ACF in time can be constructed by correlating burst pairs, each giving a value of $r(\Delta t_{jk} = t_{k}-t_{j})$ \citep{cordes+04, main+17, main+21}.  We correlate all burst pairs in their overlapping frequency channels (ie. in their overlapping on-region described in Section \ref{sec:spectra}), correcting for the effect of noise biasing the measured values of $\sigma$ (for more detail, see the appendix in \citealt{main+21}).  For Effelsberg, all pairs of bursts are correlated, while for the uGMRT only the correlation of bursts with the brightest burst are considered.

Errors are obtained through MCMC analysis.  For each pair of burst spectra $I_{j}(f), I_{k}(f)$, we measure the noise $N$ in the off-burst region, using the same weighted profile that was used to construct the spectra.  We perform 100 iterations, adding Gaussian random noise with standard deviation $N$ to each channel in $I_{j}(f), I_{k}(f)$, measure the correlation coefficient $r_{jk}$, and use the standard deviation of the resultant $r_{jk}$ values as the error on $r(\Delta t_{jk})$.

\begin{figure}
    \centering
    \includegraphics[width=1.0\columnwidth,scale=1.0, trim=0.0cm 0.0cm 0.0cm 1.5cm, clip=true]{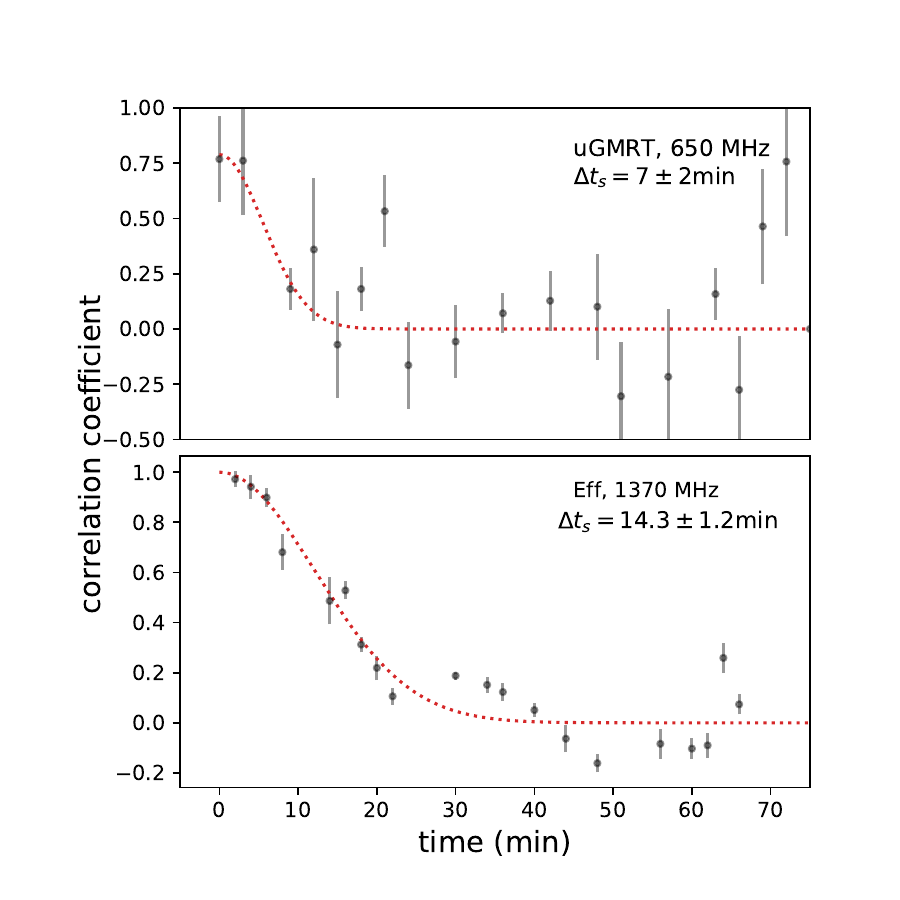} \\
    \vspace{-5mm}
    \caption{ Correlation coefficients between burst pairs at uGMRT (\textit{top}) and Effelsberg (\textit{bottom}), as described in Section~\ref{sec:scint_timescale}.  The red dotted lines show the best fit Gaussian, from which the scintillation timescale is measured. The time axis is restricted to $< 75$\,mins, and the time axes are binned to 2 and 3 minutes for visual purposes }
    \label{fig:scinttimescale}
\end{figure}

The pairwise time ACFs are shown in Figure \ref{fig:scinttimescale}.  The scintillation timescale is defined as the half width at half maximum (HWHM) of the time ACF, which we extract by fitting a Gaussian to all points $r(\Delta t_{i})$, weighted by their errors.  
We find a shorter scintillation timescale at lower frequency, with measured values of $\Delta t_{G} = 7 \pm 2$\,min, and $\Delta t_\mathrm{Eff} = 14.3 \pm 1.2$\,min, consistent with the expected linear scaling of the scintillation timescale with frequency.

\subsection{2D ACFs and power spectra}

\begin{figure*}
    \centering
    \includegraphics[width=1.0\textwidth,scale=1.0, trim=2.0cm 0.0cm 1.0cm 1.0cm, clip=true]{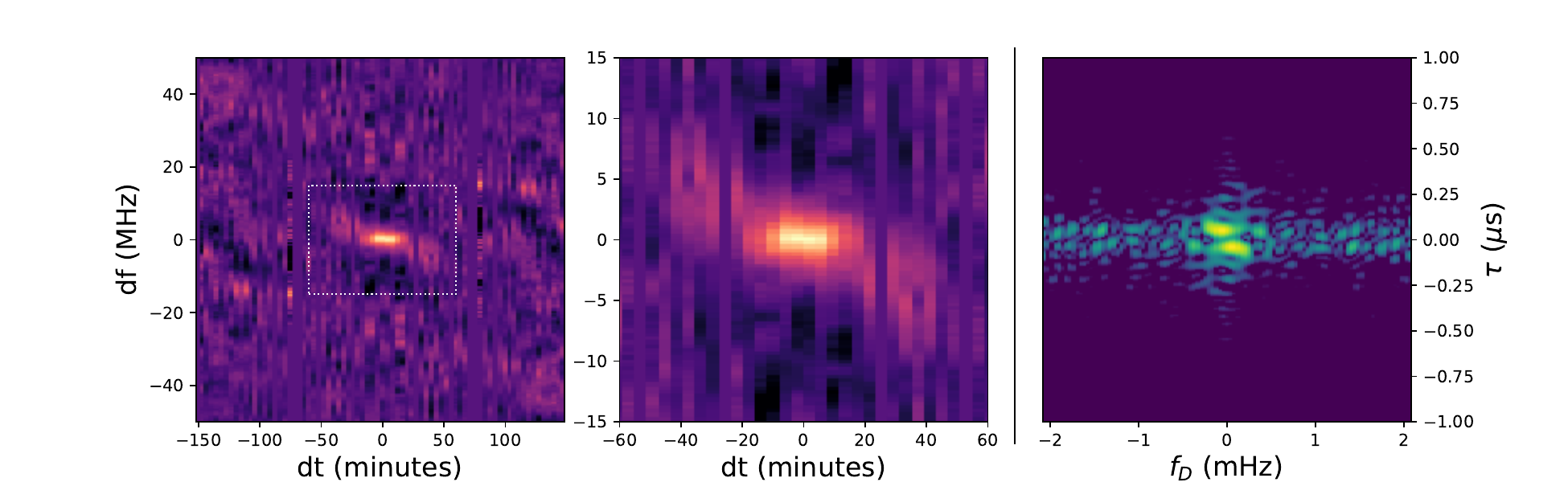} \\
    \caption{ Large-scale view (\textit{left}) and central region (\textit{middle}) of the 2D ACF of the Effelsberg bursts, and the corresponding secondary spectrum (\textit{right}), formed through a 2D FFT of the ACF.  The secondary spectrum is plotted in log scaling, and the colourbar extends two orders of magnitude. There are hints of diagonal stripes in the ACF, which result in an asymmetric power distribution in the secondary spectra}
    \label{fig:ACFs}
\end{figure*}

In the above sections, we computed the time and frequency ACFs separately, using burst pairs.  More generally, the time and frequency ACFs are cuts through the $\Delta t=0$ or $\Delta f=0$ axes of the two-dimensional ACF,
\begin{equation}
R(\Delta t, \Delta f) = \sum_{ij}\frac{ I(t_{i}, f) I(t_{j}, f + \Delta f_{j}) }{ \sigma^{2} }.
\end{equation}
In the Effelsberg data, bursts are resolved over many channels, and there are many burst pairs within the scintillation timescale.  We construct the 2D ACF by binning the frequency ACF of all burst pairs as a function of $\Delta t$ in 5\,minute bins, weighted by their errors.  The resulting ACF is shown in Figure \ref{fig:ACFs}.

A complimentary tool to the 2D ACF is the power spectrum of scintillation (often called the ``secondary spectrum''), which is the 2D Fourier transform of the ACF and has proven to be a valuable tool for studying pulsar scintillation (eg, \citealt{stinebring+01, brisken+10}).  The secondary spectrum describes the scintillation in time and frequency in terms of its conjugate variables $f_{D}$ and $\tau$, which are related to the Doppler rate and the geometric time delay between interfering images respectively.  We compute the secondary spectrum for FRB20201124A by taking the Fourier transform of the 2D ACF, and the result is shown in the right panel of Figure \ref{fig:ACFs}.

The bright central region of the 2D ACF shows the characteristic time and frequency scales of scintillation measured in the above section.  There are also hints of diagonal features, all following the same drift direction, which would imply scattering from an anisotropic screen.  The secondary spectrum, while largely featureless, shows a clear asymmetry in the power arising from the diagonal stripes in the ACF.  The bulk of power is seen at $f_{D}\sim -0.15\,$mHz, $\tau \sim 0.1 \mu$s, arising from the reciprocal of $t_{s}$ and $\nu_{s}$.

\section{RESULTS}
\label{sec:discussion}

\subsection{The low scattering of {\frb}}
\label{sec:underscattered}

From our measurements of $\nu_{s}$, we can estimate the scattering time $\tau_{s} \approx 1 / 2\pi\nu_{s}$.  The standard measure of $\tau_{s}$ is scaled to 1\,GHz;  we measure $\tau_{1\rm GHz}=0.31\pm0.06\,\mu$s, where the mean value comes from the best-fit power-law of $\nu_{s}$ across frequency in Figure \ref{fig:scintbandwidth}, and the errors are conservatively estimated by the upper and lower limits obtained by extrapolating a $\nu_{\rm s} \propto \nu^{4}$ from uGMRT or from Effelsberg. 

\begin{figure}
    \centering
    \includegraphics[width=1.0\columnwidth,scale=1.0, trim=0.0cm 0.0cm 0.0cm 0.0cm, clip=true]{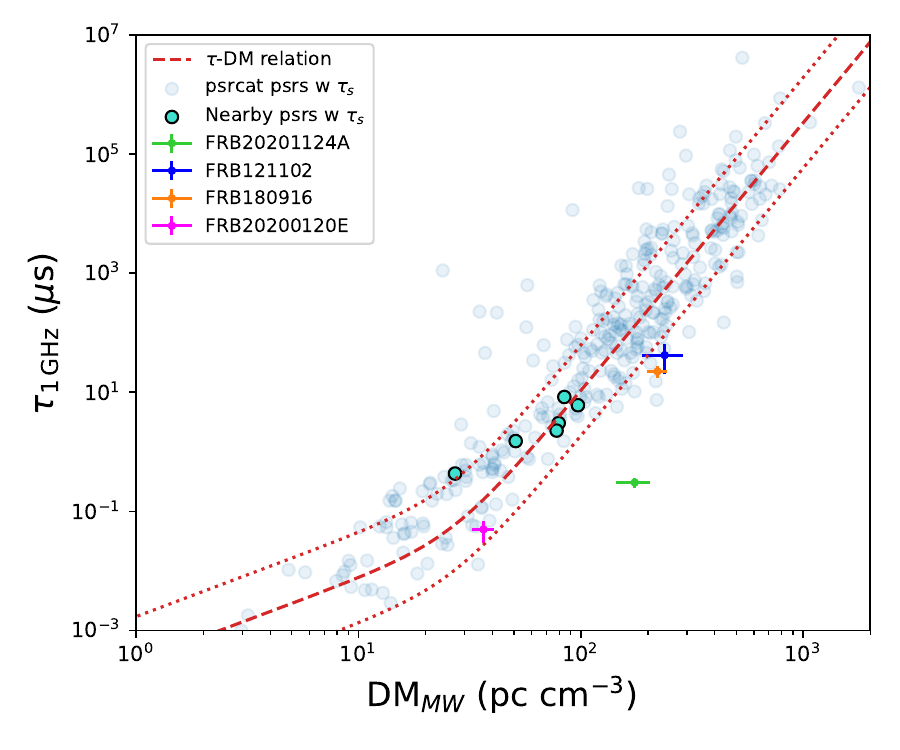} \\
    \vspace{-3mm}
    \caption{ Measurement of $\tau_{\rm 1 GHz}$ for FRB20201124a plotted on the $\tau-$DM relation.  The red dashed and dotted lines show the $\tau-$DM relation and $1\sigma$ scatter from \protect{\citet{cordes+16}}. The range of DM values for the FRBs in the plot span the NE2001 and YMW16 values for the MW component of the DM. }
    \label{fig:taudm}
\end{figure}

In Figure \ref{fig:taudm}, we plot the $\tau$-DM relationship for all known pulsars with scattering values from 
PSRCAT\footnote{\url{http://www.atnf.csiro.au/research/pulsar/psrcat}} \citep{manchester+05}, 
as well as measurements of scattering for {\rone} and {\rthree} from \citet{ocker+21}, and {\rGC} from \citet{nimmo+21b}. Since the observed scintillation of {\frb} is most likely from the MW, we use only the MW contribution of the DM (denoted as DM$_{\rm MW}$), estimated using NE2001 \citep{cordes+02} or the YMW16 \citep{yao+17} Galactic electron models.

{\frb} appears significantly underscattered compared to other sources of similar 
DM$_{\rm MW}$, including {\rone}, {\rthree} and {\rGC}.  In addition, {\frb} is underscattered 
compared to nearby pulsars; there are six pulsars within 20 degrees with scattering 
measurements from PSRCAT, all of which have a larger $\tau_{1 \rm GHz}$ despite being seen 
through less of the MW.

\subsection{Predictions from electron-density models}
\label{sec:NE2001}
We wish also to compare the scattering of {\frb} to the predictions of galactic electron-density models, including NE2001 \citep{cordes+02} and YMW16 \citep{yao+17}.

\begin{figure}
    \centering
    \includegraphics[width=1.0\columnwidth,scale=1.0, trim=0.0cm 0.0cm 0.0cm 0.0cm, clip=true]{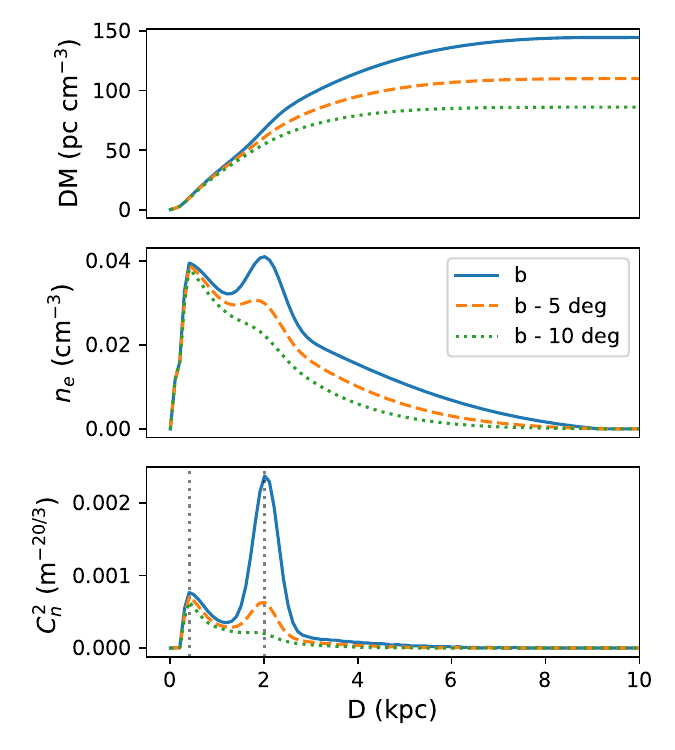} \\
    \vspace{-3mm}
    \caption{ Model predictions from NE2001 for the total DM (\textit{top}), the electron density (\textit{middle}), and the scattering strength (\textit{bottom}) along the line-of-sight towards {\frb}. The orange and green curves show model predictions a further $5^{\circ}$ and $10^{\circ}$ below the Galactic plane. The expected source of scattering towards \frb{} occurs in a spiral arm at $\sim 2\,$kpc, which is missed at lower Galactic latitudes.} 
    \label{fig:NE2001}
\end{figure}

To estimate $\tau$ from NE2001, we use the predicted value of extragalactic broadening angle $\theta_{x}$ along with a thin screen model, as is done in \citet{ocker+21}, where the equation for $\tau$ arising from a thin screen is
\begin{equation}
\tau = \frac{\theta^{2}}{8\rm{ln(2)}c} \frac{D d_{l}}{D - d_{l}} \approx \frac{\theta^{2} d_{l}}{8\rm{ln(2)}c}, \text{when }D \gg d_{l}.
\end{equation}
 To estimate the screen distance, we plot the NE2001 model values of DM, $n_{e}$, and $C_{n}^{2}$ as a function of distance towards {\frb} in Figure \ref{fig:NE2001}.  $C_{n}^{2}$ (roughly speaking a measure of the strength of scattering in a region) peaks at $d_{l} \approx 2.0$\,kpc. 

\begin{figure*}
    \centering
    \includegraphics[width=0.7\textwidth,scale=1.0, trim=0.5cm 1.5cm 0.5cm 2.5cm, clip=true]{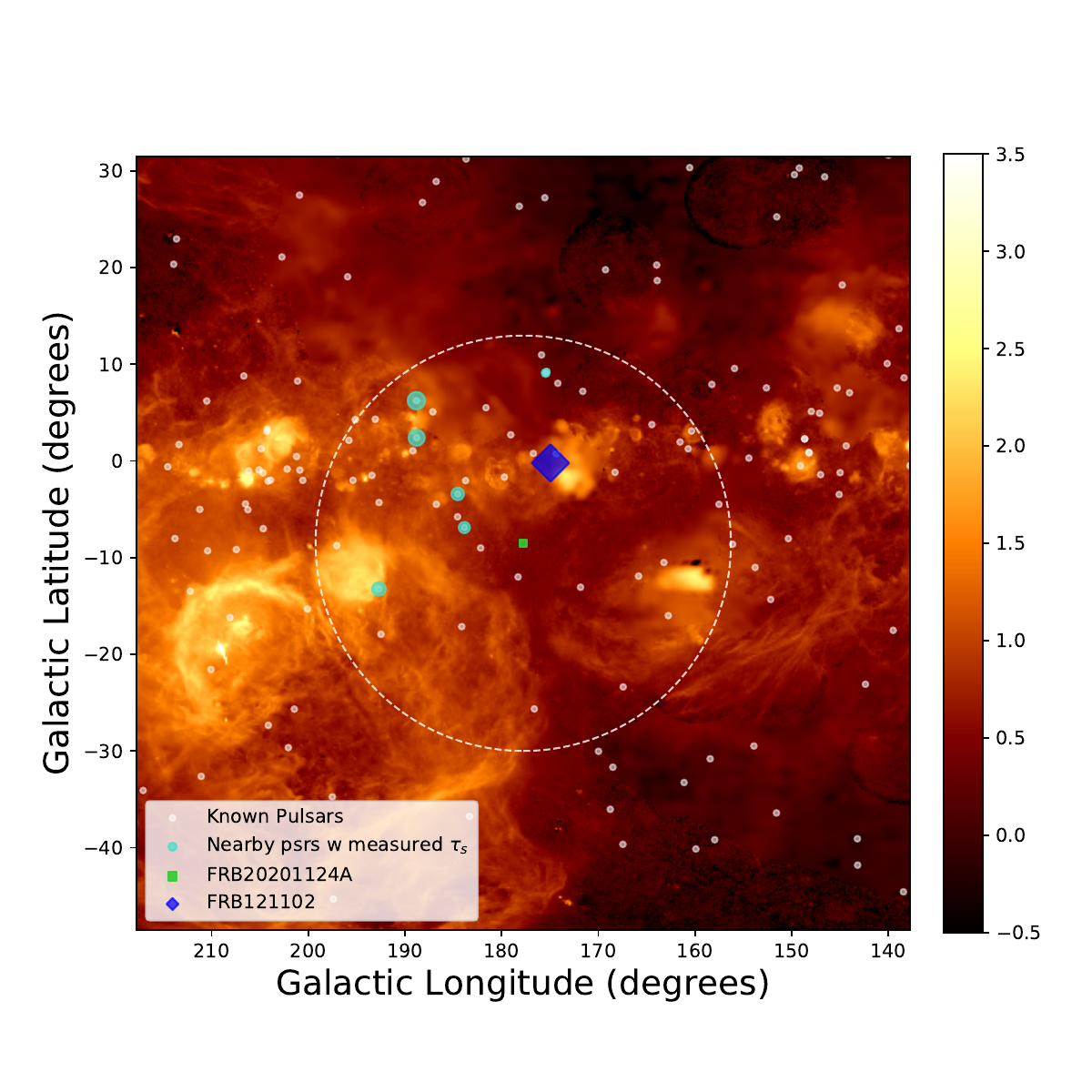} \\
    \caption{H$\alpha$ map surrounding the line of sight of {\frb}, from \protect{\citet{finkbeiner03}}.  The colorbar is logarithmic, in units of Rayleighs. The dashed circle is a radius of 20$^{\circ}$, chosen to contain pulsars on nearby lines-of-sight with known scattering measurements. The size of the coloured points represents measured values of $\tau_{1\,\rm GHz}$ in Figure \ref{fig:taudm}, in a square root scaling.}
    \label{fig:halpha}
\end{figure*}

The model $\theta_{x} \approx 3.336\,$mas, combined with $d_{l}\approx 2\,$kpc gives $\tau_{1 \rm GHz, NE2001} \approx 9.7\mu$s, a factor of $\approx 30$ greater than our measured value. Angular broadening is not output in YMW16, but they predict an asymptotic value of $\tau_{1 \rm GHz, YMW16} \approx 0.4\,$ms, even more discrepant with our values.
By any metric, {\frb} is significantly underscattered, or has a much lower value of ${\rm DM}_{\rm MW}$ than predicted from NE2001 or YMW16.  
To investigate why {\frb} may be underscattered, we look at its surroundings using the all-sky H$\alpha$ map from \citet{finkbeiner03} in Figure \ref{fig:halpha}. H$\alpha$ emission is a tracer of ionized Hydrogen, and is proportional to the scattering measure.  {\frb} is off of the Galactic plane, seen through a comparatively quiet region, with little surrounding H$\alpha$ emission suggestive of any particular scattering regions.  We note, however, that the angular scales of relevant structures for scattering ($\sim$ mas) are far below the pixel resolution of the map.

\subsection{Low Scattering Lines of Sight}
\label{sec:VLBI}

Galactic electron density models like NE2001 are very useful for estimating 
typical values for DM and scattering along any given line of sight.  However, 
the scattering properties can vary significantly on small angular scales, and 
this cannot be modelled for every line of sight.  Additionally, {\frb} is extragalactic, lying in the direction of the galactic anticenter, where there are comparatively few scattering measurements. 
One way to get a sense for 
the variability of scattering is to use a large multi-frequency catalog of 
AGN core sizes.  The catalog of \citet{pushkarev+15} gives VLBI measured 
sizes of AGN cores for thousands of sources at 2.3 and 8.4~GHz (and other 
frequencies up to 43~GHz for some sources).  VLBI measurements of AGN 
core sizes are excellent probes of increased scattering in the galactic plane where 
scattering sizes can be several milliarcseconds even at 2.3~GHz.  However, 
the intrinsic sizes of these sources are $\approx1$~mas at 2.3~GHz, so it is 
difficult to extract scattering sizes much smaller than this.  Higher 
frequency measurements of the AGN sizes can be used to determine the 
unscattered size of the AGN core, but the intrinisc size of these 
cores also changes with frequency \citep{pushkarev+15}.  
For the four sources within five degrees of {\frb}, we estimate 2.3~GHz scattering 
sizes are between $\approx 0-1.5$~mas, which would correspond to 
$\approx 0-8$~mas at 1~GHz.  Unfortunately, the fundamental limiting resolution 
of the VLBI observations, the frequency-dependent intrinsic size of the sources, 
and the small amount of scattering involved prevents us from setting more useful 
limits on the angular broadening seen toward these sources.  Because these 
problems will apply to all VLBI observations of small scattering sizes, FRBs may 
be one of the only ways to measure low scattering lines of sight in regions of the 
sky with few pulsars.

\subsection{Scintillation Velocity}

For a thin screen, the scintillation velocity $V_\mathrm{ISS}$ can be derived from $\nu_{s}, t_{s}$ as
\vspace{-1.5mm}
\begin{equation}
V_\mathrm{ISS} = A_\mathrm{ISS} \frac{\sqrt{D \nu_{s} }}{f t_{s}}~\textrm{km~s}^{-1}, \quad  A_{ISS} \approx 27800 \sqrt{\frac{2(1-s)}{s}},
\end{equation}
with $D$ in kpc, $\nu_{s}$ in MHz, $f$ in GHz, $t_{s}$ in seconds,
$s = 1 - d_{l}/D$, and the prefactor of $A_\mathrm{ISS}$ depends on the exact properties of the scattering screen, this particular prefactor coming from an isotropic Kolmogorov screen \citep{cordes+98}.
The scintillation velocity depends on the relative velocities of the source, Earth, and screen,
\vspace{-1.5mm}
\begin{equation}
\textbf{V}_\mathrm{ISS} = (1-s) \textbf{V} + s \textbf{v}_{\earth} - \textbf{v}_{l}.
\end{equation}
For an extragalactic source, $D \gg d_{l}, s \rightarrow 1$ and the above equations simplify, with the source distance and velocity becoming negligible:
\vspace{-1.5mm}
\begin{equation}
    V_\mathrm{ISS} \approx 27800~\textrm{km~s}^{-1} \frac{\sqrt{2 \nu_{s} d_{l} }}{f t_{s}} \approx |\textbf{v}_{\earth} - \textbf{v}_{l}|_{||}.
\end{equation}
Using the measured $\nu_{s}, t_{s}$ from Effelsberg or uGMRT, we obtain $V_\mathrm{ISS} = (59\pm7) \sqrt{d_{l}/2\,\rm{kpc}}~{\rm km~s}^{-1}$ or $V_\mathrm{ISS} = (64\pm20) \sqrt{d_{l}/2\,\rm{kpc}}~{\rm km~s}^{-1}$, respectively.  The value of $V_\mathrm{ISS}$ is larger than the velocity of the Earth, and would imply a transverse velocity of $\sim30-40~{\rm km~s}^{-1}$ of the dominant scattering screen.  

\begin{figure}
    \centering
    \includegraphics[width=1.0\columnwidth,scale=1.0, trim=0.0cm 0cm 0.0cm 0cm, clip=true]{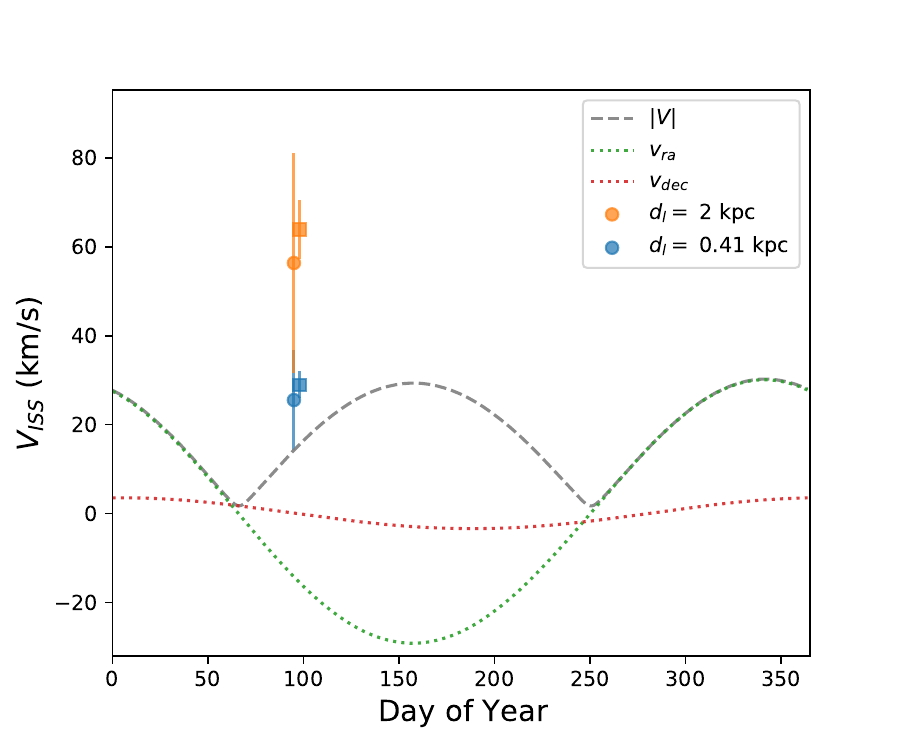} \\
    \caption{Scintillation velocities at two different screen distances, and the velocity of the Earth throughout the year.  The square and circle shows the Effelsberg and uGMRT measurement, respectively. While a single point is not constraining, an annual variation could be modelled to measure screen parameters. }
    \label{fig:scintvelocity}
\end{figure}

From Fig \ref{fig:NE2001}, there is an expected scattering contribution in the environment closer to Earth, with a peak at $\sim 0.4\,$kpc.  If scattering is dominated by this region (either seen through a hole or `quiet' patch of the spiral arm), then the scintillation velocity is of order $\sim30-40~{\rm km~s}^{-1}$, and could be naturally explained through the velocity of the Earth and a small screen velocity.  This would also explain the dearth of scattering described in \ref{sec:underscattered} - the NE2001 model prediction $10^{\circ}$ lower in Galactic latitude is dominated by the local peak at $\sim400\,$pc, and predicts a scattering time of $\tau \approx 0.47\,\mu$s, much closer to our measurement.
The scintillation velocities for both screen distances are shown in Figure~\ref{fig:scintvelocity}, along with the velocity of the Earth on the plane of the sky towards \frb{}.


The geometry and distance of the scattering screen could be obtained through annual variation of the scintillation timescale.  These techniques have been successfully applied to variable scintillation timescales of extragalactic sources seen and modelled in intraday variable quasars, often finding screens very close to the Earth \citep{bignall+03, walker+09}. Similar techniques have been used in pulsar scintillation to constrain scattering screen distances, geometries, and velocities through annual variations of scintillation arcs \citep{rickett+14, reardon+20, main+20}, or using inter-station time delays or VLBI \citep{brisken+10, simard+19}.

\subsection{Scintillation / scattering in host}

The above section focused on scintillation in the MW.  However, many FRBs exhibit scattering which is likely from their host galaxy (eg. \citealt{masui+15}), or exist in local extreme magnetoionic regions which may be conducive to propagation effects \citep{michilli+18, hessels+19, hilmarsson+21}.  If one could measure the scintillation timescale in the host (or local environment), then one could probe the velocity of the FRB and constrain the emission regions \citep{simard+20}.  If the active periods in repeating FRBs are caused by an orbit \citep{16DaiOrbit, 20IokaOrbit}, then one could see a periodic modulation of the host scintillation timescale arising from orbital motion - exactly the same principle is used to constrain binary parameters of pulsars \citep{lyne84, rickett+14, reardon+20}.

In the regime where one sees a scattering tail (likely from the FRB's host galaxy or local environment) rather than an observable scintillation pattern (likely from the MW), then using baseband data, one could still measure the scintillation timescale associated with the screen causing the temporal scattering.  The scattering tail would coherently correlate over the scintillation timescale, which could be determined through the coherent correlation of scattered bursts nearby in time.
A proof of concept was shown for giant pulses in B1957+20, in which the scattering tails of nearby pulse pairs could be used to coherently de-scatter each other, and the decorrelation timescale matched the scintillation timescale measured using traditional methods \citep{main+17}.

\subsection{Host screen constraint from 2-screen model}

 In \citet{masui+15}, FRB 110523 was found to exhibit both scattering and scintillation, with scattering likely from the FRB's host galaxy, and the scintillation pattern in frequency from the MW, which persisted throughout the scattering tail.  They argued that for the scintillation to appear constant throughout the scattering tail, the scattering disk in the host galaxy must be `unresolved' by the MW scattering screen (this argument is also outlined in \citealt{cordes+19}).  We employ a similar argument below, to investigate a potential 2-screen model for \frb{}.

The effective resolution of the MW screen (i.e. the angular distance at the source over which the scintillation pattern would change by $\pi$ radians) is
\begin{equation}
\Theta_\mathrm{res} = \lambda/D = \frac{\lambda}{d_{l,\rm MW} \theta_{\rm MW}} = \lambda \left(\frac{\pi \nu_{\rm s, MW}}{c d_{l,\rm MW}} \right)^{1/2},
\end{equation}
and the angular size of the scattering screen in the host is 
\begin{equation}
\theta_{\rm h} = \left(2 c \tau_{\rm h} \frac{D-d_{l,\rm h}}{D d_{l,\rm h}}  \right)^{1/2} \approx \frac{(2c\tau_{\rm h} d_{ls, \rm h})^{1/2}}{D},
\end{equation}
where the subscripts $\rm MW$ and $\rm h$ denote scattering in the MW or host galaxy, respectively, and $d_{ls, \rm h} \equiv D-d_{l,\rm h}$.

{\frb} has been localized to a host galaxy, which has a spectroscopic redshift of $z=0.098\pm0.002$ \citep{redshiftatel, fong+21, ravi+21}, corresponding to an angular diameter distance of $D \approx 374$\,Mpc (using approximate values of $\Omega_{m}=0.3, \Omega_{\Lambda}=0.7, H_{0}=70$).  The distance between the FRB source and the screen is $d_{ls,\rm h} \equiv D - d_{l, \rm h}$.  From the NE2001 estimate from Section~\ref{sec:NE2001}, we use $\equiv d_{l,MW} \sim 2\,$kpc.  In P-I, the brightest uGMRT burst (our reference uGMRT burst in Figure \ref{fig:dynamic_spectra}) appears to have an exponential tail at 550-600 MHz - if this is interpreted as scattering, rather than a shape coincidence from the intrinsic emission, then the scattering in the host is constrained to $\tau_{h} \approx 11\,$ms.

For scintillation to occur, the inequality $\Theta_\mathrm{res} > \theta_{\rm h}$ must be met.  Combining equations 6 and 7,
\begin{equation}
    d_{ls,\rm h} \lesssim \frac{\pi \lambda^{2} \nu_{\rm s, MW} D^{2} }{2 c^{2} \tau_{\rm h} d_{\rm l,MW}}.
\end{equation}
Using the lower uGMRT value of $\nu_{s}\approx 0.1\,$MHz gives $d_{ls,\rm h} \lesssim 3\,$kpc, constraining the scattering to occur within the host galaxy.
If scintillation is still resolved through the scattering tail at lower frequencies, this constraint will become much stronger.  If $\tau \propto \lambda^{4}$, then the inequality scales as 
$\lambda^{-6}$.
If one could measure the transition frequency where the MW scintillation is quenched throughout the scattering tail arising in the host, and there was a measure of $d_{\rm l,MW}$, then one could determine $d_{ls,\rm h}$.

\section{SUMMARY AND CONCLUSIONS}
\label{sec:conclusion}

We have measured a scintillation timescale of {\frb}, and the inferred scintillation velocity suggests a velocity of 10s of km/s of the intervening scattering screen, or that the scattering screen is much closer to the Earth than at $\sim 2\,$kpc which is the expected peak of scattering from the NE2001 model.  While from a single measurement the scattering screen distance and geometry are still unknown, this measurement serves as a proof-of-concept for using scintillation to probe dynamics of FRBs.  If this, and other FRBs remain similarly active over time, then one can model annual variations of scintillation timescales to obtain screen distances and geometries, methods which have been used successfully in intraday variable quasars and in pulsar scintillation.  For single observations, similar constraints could be obtained through interstation time delays, or through VLBI.

While we do not know if {\frb} will remain as active as it is currently, we know that other sources like {\rone} and {\rthree} have remained steadily active over several years, where sensitive observations have revealed many burst detections over short periods of time (with the caveat that these sources are only visible in periodic active windows, \citealt{chime20_periodicity, rajwade+20}).  While we were fortunate that {\frb} had comparatively low scattering in the MW, resulting in easily resolvable scintillation, some existing data of other repeating FRBs may already exist which is suitable for measuring scintillation timescales, including high frequency observations, or data taken for VLBI localization, where baseband data can be re-reduced to any frequency channelization.  {\rGC} is another source for which a scintillation timescale may be derived; the scintillation bandwidth is comparably large to {\frb}, and \citet{nimmo+21b} find that the spectra of two bursts separated by $\sim 5$\,minutes are still partially correlated.

We find {\frb} to be significantly underscattered compared to NE2001 and YMW16 predictions, 
and to nearby pulsars, despite being seen through the entire MW.  This suggests 
agreement with \citet{ocker+21} that the MW halo has an insignificant contribution 
to scattering.  While the scattering of {\rone} and {\rthree} match with the $\tau$-DM 
relation, they are both seen through the Galactic plane, while {\frb} is $8^{\circ}$ 
below the plane in the direction of the Galactic anticenter, in a region mostly free of H$\alpha$ 
emission.  For \frb{} to match the $\tau$-DM relation would imply DM$_{\rm MW}$ is lower by a factor of $\sim 2-3$.

The underscattering combined with the measured scintillation velocity are consistent with NE2001 if the scattering occurs in the local environment $\sim 400\,$pc, rather than in a spiral arm at $2\,$kpc (although more observations are needed to say this conclusively).  
These results suggest that FRBs are a useful tool in understanding scattering of 
the MW: as FRBs seem to sample the sky uniformly \citep{chime2021_gallat}, they will 
be of particular use in constraining regions relatively free of pulsars, including the 
Galactic anticenter and regions of high Galactic latitude.  Since these regions 
also tend to have low levels of scattering that are difficult to measure from angular 
broadening of AGN cores, FRBs will be a very useful tool in mapping out low scattering 
lines of sight.  Additionally, as discussed in \citet{walker+20}, the detection of many FRBs across the sky will allow for a reconstruction of DM$_{\rm MW}$.  Constraints on DM$_{\rm MW}$ in combination with the possible constraints on scattering and screen distances will be a valuable tool in constructing accurate galactic electron models.

\section{ACKNOWLEDGEMENTS}
We thank the reviewer for many useful comments which improved this paper. 
RAM thanks Charles Walker for useful comments, and discussion about determining MW DMs using FRBs.  We thank the staff of the GMRT and Effelsberg who have made these observations possible. The GMRT is run by the National Centre for Radio Astrophysics of the Tata Institute of Fundamental Research.  The Effelsberg 100-m telescope is operated by the Max-Planck-Institut f{\"u}r Radioastronomie.  VRM acknowledges the support of the Department of Atomic Energy, Government of India, under project no. 12-R\&D-TFR-5.02-0700. We acknowledge use of the CHIME/FRB Public Database, provided at \url{https://www.chime-frb.ca/} by the CHIME/FRB Collaboration. 
Part of this research was carried out at the Jet Propulsion Laboratory, California Institute of Technology, under a contract with the National Aeronautics and Space Administration.

\section{DATA AVAILABILITY}
The data underlying this article will be shared on reasonable request to the corresponding authors.

\bibliographystyle{mnras}
\bibliography{frb20201124a}

\end{document}